\documentclass[conference]{IEEEtran}
\usepackage[nolist]{acronym}
\usepackage{graphics}
\usepackage{amsmath}

\usepackage{mathtools}
\usepackage{amsmath}
\usepackage{algorithm} 
\usepackage{algpseudocode}

\usepackage{tikz}
\usetikzlibrary{shapes,shadows,arrows}
\tikzstyle{decision} = [diamond, draw, fill= blue!50]
\tikzstyle{line} = [draw, -latex']
\tikzstyle{elli} = [draw, ellipse, fill=red!50, minimum height = 8mm]
\tikzstyle{block} = [draw, rectangle, fill= blue!50, text width=8em, text centered, minimum height = 15mm, node distance=5em]
\tikzstyle{line} = [draw,-latex']
\usepackage{amsmath}
\usepackage{tikz}
\usepackage{mathdots}
\usepackage{yhmath}
\usepackage{cancel}
\usepackage{color}
\usepackage{siunitx}
\usepackage{array}
\usepackage{multirow}
\usepackage{amssymb}
\usepackage{gensymb}
\usepackage{tabularx}
\usepackage{extarrows}
\usepackage{booktabs}
\usetikzlibrary{fadings}
\usetikzlibrary{patterns}
\usetikzlibrary{shadows.blur}
\usetikzlibrary{shapes}
\usepackage{subcaption}
\usepackage{amsmath, amssymb, bm}
\usepackage{caption}
\usepackage{subcaption}

\usepackage{xcolor}
\usepackage{soul}
\usepackage[top=0.8in, left=0.65in, right=0.65in, bottom=1.1in]{geometry}
\renewcommand{\baselinestretch}{.950}

\begin{document}
\pagenumbering{arabic}

\title{Energy Consumption Reduction for UAV Trajectory Training : A Transfer Learning Approach}




\vspace{-6mm}

\author{\IEEEauthorblockN{Chenrui Sun\IEEEauthorrefmark{1}, Swarna Bindu Chetty\IEEEauthorrefmark{1}, Gianluca Fontanesi\IEEEauthorrefmark{2}, Jie Zhang\IEEEauthorrefmark{1}, Amirhossein Mohajerzadeh\IEEEauthorrefmark{3},\\ David Grace\IEEEauthorrefmark{1} and Hamed Ahmadi\IEEEauthorrefmark{1}}
\\
\IEEEauthorrefmark{1}School of Physics Engineering and Technology, University of York, United Kingdom\\
\IEEEauthorrefmark{2}Interdisciplinary Centre for Security, Reliability, and Trust (SnT), Luxembourg
\IEEEauthorrefmark{3}Sohar University, Oman\\

Email: chenrui.sun@york.ac.uk,
swarna.chetty@york.ac.uk,
gianluca.fontanesi@ieee.org, \\ jie.zhang@york.ac.uk, mohajerzadeh@um.ac.ir,  david.grace@york.ac.uk, hamed.ahmadi@york.ac.uk
}

\maketitle
\vspace{-4mm}

\begin{abstract}


The advent of 6G technology demands flexible, scalable wireless architectures to support ultra-low latency, high connectivity, and high device density. The \ac{O-RAN} framework, with its open interfaces and virtualized functions, provides a promising foundation for such architectures. However, traditional fixed base stations alone are not sufficient to fully capitalize on the benefits of O-RAN due to their limited flexibility in responding to dynamic network demands. The integration of \acp{UAV} as mobile RUs within the O-RAN architecture offers a solution by leveraging the flexibility of drones to dynamically extend coverage. However, \ac{UAV} operating in diverse environments requires frequent retraining, leading to significant energy waste. We proposed transfer learning based on Dueling \ac{DDQN} with multi-step learning, which significantly reduces the training time and energy consumption required for \acp{UAV} to adapt to new environments.We designed simulation environments and conducted ray tracing experiments using Wireless InSite with real-world map data. In the two simulated environments, training energy consumption was reduced by 30.52\% and 58.51\%, respectively. Furthermore, tests on real-world maps of Ottawa and Rosslyn showed energy reductions of 44.85\% and 36.97\%, respectively.

\end{abstract}
\IEEEpeerreviewmaketitle

\acresetall

\vspace{-2mm}

\section{Introduction}

The shift to \ac{6G} networks brings unprecedented demands for ultra-low latency, extreme data rates, and support for over 10 million devices per square kilometer, including IoT systems, autonomous vehicles, and smart city infrastructure. Meeting these diverse Quality of Service (QoS) requirements while ensuring energy efficiency, is challenging for traditional architectures \cite{polese2023understanding}. These systems, often controlled by single vendors, lack the flexibility, scalability, and open interfaces necessary to dynamically allocate resources and optimize energy consumption—key for fulfilling \ac{6G}’s demanding Key \ac{KPIs} such as sub-millisecond latency and data rates exceeding 1 Tbps. \ac{O-RAN} provides a flexible, scalable solution for 6G, enabling interoperability through open interfaces and the disaggregation of components like \ac{O-CU}, \ac{O-DU} and \ac{O-RU}. Virtualizing CUs and DUs in the O-Cloud boosts flexibility and enhances energy efficiency, while the modular approach allows for more efficient scaling physical infrastructure. \ac{RIC} enable real-time and long-term AI/ML-based optimizations. The \ac{Near-RT RIC} manages real-time traffic adjustments, while the \ac{Non-RT RIC} focuses on longer-term optimizations like capacity planning \cite{liang2024energy}. These AI-driven controls allow the network to dynamically adapt, optimize resources, and minimize energy use, making O-RAN essential for the energy-efficient, high-performance demands of 6G.

Building upon this foundation, the integration of \ac{UAV} with \ac{O-RAN} architecture further enhances network capabilities by providing dynamic coverage, three-dimensional network optimization, and temporary capacity enhancement~\cite{sun2024advancing}. Within this framework, \acp{UAV} can serve as mobile and flexible \acp{O-RU}, offering distinct advantages. They enable on-demand coverage extension and capacity enhancement, allowing for rapid repositioning to address traffic spikes or temporary outages. Furthermore, the ability to adjust \ac{UAV} altitude optimizes signal propagation and minimizes interference, leading to improved overall network performance. The rApps/xApps in \ac{RIC} can optimize \ac{UAV} trajectories and resource utilization, adapting to changing network conditions and user demands. This dynamic control allows for more efficient network operations, especially given the variable nature of \ac{UAV} deployments and fluctuating user traffic~\cite{cumino2024enhancing}. A first tentative to integrate a flexible multi-\acp{UAV} system with the \ac{O-RAN} architecture, referred to as U-ORAN, has been presented in \cite{pham2022ran}. This approach utilizes multi-UAV trajectories and task offloading through  multi-agent \ac{RL} and online learning methodologies.

However, a significant challenge in using \ac{RL} for \ac{UAV} trajectory planning in O-RAN contexts is the need for extensive retraining when \acp{UAV} encounter new environments. This retraining process consumes substantial computational energy, reducing the overall efficiency and delaying the deployment of \acp{UAV} in new areas. Therefore, addressing the retraining issue is crucial for enhancing the practical deployment of \ac{UAV}-based communication systems within \ac{O-RAN} architectures. Previous works as \cite{Sun_wincom24} address the retraining problem using \ac{TL}, successfully reducing retraining time in \ac{UAV}-assisted emergency communication and tested in a simulated environment. However, this approach does not take into account the energy consumption associated with training. Another significant limitation in most prior works is that they focus on fixed target locations and single tasks, neglecting the complexity of dynamic users and target locations. Realistically, the target of a \ac{UAV} is more likely to involve reaching any location within a defined area rather than a single specific point. Additionally, existing O-RAN architecture solutions and \ac{TL} methods have primarily been tested in simulated environments without validation in real city maps to prove their reliability. This research investigates how \ac{TL} can reduce training energy consumption of \ac{DRL} \ac{UAV} trajectory and enhance performance within the O-RAN architecture. By incorporating ray-tracing techniques with real maps and materials, we able to validate the approach in real city maps scenarios.

\begin{figure}[ht]
\vspace{-2mm}

    \centering
    \includegraphics[width=0.85\columnwidth]{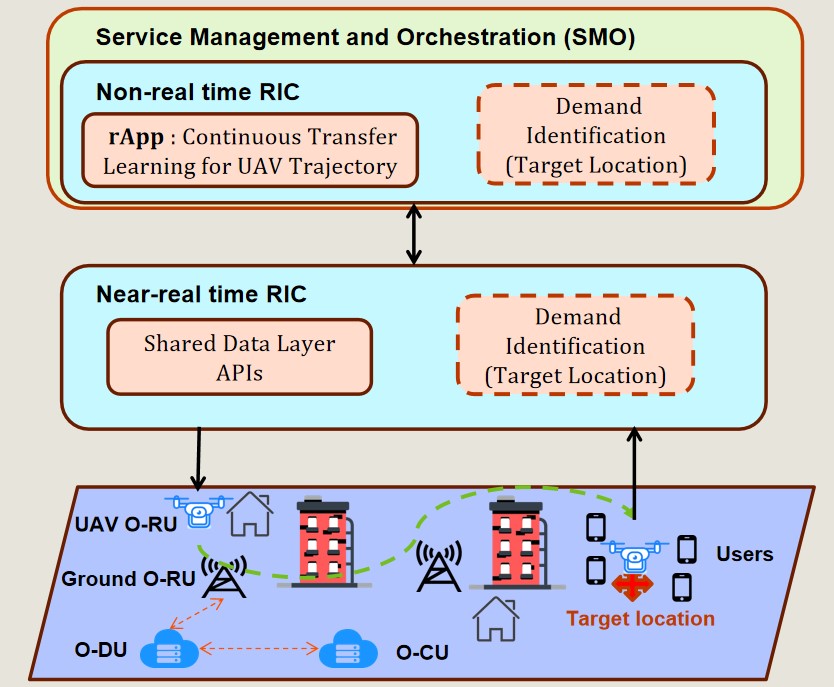}
    \caption{UAV O-RAN system}
    \label{fig:structure}
\vspace{-6mm}

\end{figure}

\vspace{-0.07in}
\subsection{Contributions}

In this paper, we propose a novel approach where \acp{UAV} functions as mobile \ac{O-RU} within the O-RAN architecture, leveraging the flexibility of \ac{UAV} and the programmability of O-RAN to create a highly adaptable network infrastructure. Our solution addresses the challenge of optimizing \ac{UAV} trajectories for coverage across areas, considering the best target location based on user distributions. We are the first to propose an approach that accounts for dynamic users and shifting target locations within the target area. To achieve this, we employ Dueling \ac{DDQN} with multi-step learning for trajectory optimization. Furthermore, we implement \ac{TL} to significantly reduce the training time and energy consumption associated with adapting to new environments, enhancing the system's adaptability and reducing computational overhead. To rigorously test and validate our transfer trajectory system, we have created both simulation environments and ray tracing environments using Wireless Insite. We are the first to apply \ac{TL} for \ac{UAV} trajectory optimization in a ray tracing environment, combining with real-world map data to assess the system's performance under practical conditions.
\vspace{-2mm}

\section{System Model and Problem Formulation}

\subsection{System Model}
 
The proposed O-RAN-based \ac{UAV} system architecture employs a three-tiered approach shown as Fig. \ref{fig:structure}, with a \ac{Non-RT RIC} hosting a \ac{CTL} for \ac{UAV} trajectory rApp for strategic planning, a \ac{Near-RT RIC} for tactical control, and \ac{UAV} \ac{O-RU} for physical operations. The Demand Identification module for target location flexibly operates in either the \ac{Non-RT RIC} or \ac{Near-RT RIC}, depending on latency requirements. The Shared Data Layer acts as a central repository and communication hub, facilitating efficient data exchange between different components of the system and enabling real-time access to critical information for both strategic planning and operational decision-making across the O-RAN \ac{UAV} architecture. When entering a new environment, the \ac{UAV} downloads a pre-trained base model for \ac{TL} and the user-determined destination, enabling adaptive performance. \ac{UAV} needs to navigate to the designated target location to provide service while maintaining high-quality ground links and avoiding outage areas. After completing the new training, the \ac{UAV} uploads information about the environment and the trained model. 

We model the urban landscape as a square region of $D \times D$ km$^2$, populated with $N$ buildings. Our channel model incorporates path loss and antenna gain, crucial for accurately representing the wireless environment. The path loss is typically classified into two categories: Line-of-Sight (LoS) and Non-Line-of-Sight (NLoS). The LoS path loss model is given by $l_L(d) = X_L \times d^{-\alpha_L}$, while the NLoS model is expressed as $l_{NL}(d) = X_{NL} \times d^{-\alpha_{NL}}$\cite{fontanesi2020outage}. The antenna gain is modeled using a 3D pattern that accounts for both horizontal and vertical components, expressed as $G(\theta, \phi) = G_{\text{max}} - \min\left\{-\left[A_H(\phi) + A_V(\theta)\right], A_m\right\}$, where $G_{\text{max}}$ is the maximum gain, $A_H(\phi)$ and $A_V(\theta)$ represent the horizontal and vertical antenna patterns, respectively, and $A_m$ is the maximum attenuation. Fixed ground \ac{O-RU} are deployed throughout the urban area. Each O-RU is equipped with a Uniform Planar Array (UPA) antenna with dimensions of $1 (horizontal) \times 8 (vertical) $ elements. In addition to the simulated environment, we utilized Wireless Insite, a commercial ray-tracer by Remcom \cite{remcom2023wireless}, for more realistic testing using a real city map.

Within this O-RAN structure, a single \ac{UAV} carries a O-RU with capabilities similar to the ground O-RU. Its initial position $p_s$ $(x_{\text{UAV}}, y_{\text{UAV}}, h_{\text{UAV}})$ is randomly set within a designated launch area $\mathcal{A}_{\text{launch}} \subset \mathbb{R}^3$, subject to height constraints $h_{\text{min}} \leq h_{\text{UAV}} \leq h_{\text{max}}$. To model user demand, we consider $M = 40$ users randomly distributed within a specific target area $\mathcal{A}_{\text{target}} \subset \mathbb{R}^2$, following the distribution $(x_i, y_i) \sim \text{Uniform}(\mathcal{A}_{\text{target}})$, for $i = 1, 2, ..., M$. This distribution reflects a scenario where service is needed in a particular region of the urban environment. The trajectory optimization involves planning the \ac{UAV}'s path from $p_s$ to target position \( p^*_i \) : \( \{p^*_1, p^*_2, \dots, p^*_M\} \) represent multiple dynamic locations, each corresponding to different user distributions in the target area. This process considers various parameters, including a constant maximum speed $V_{max}$ and mission duration $N$. The trajectory, represented as a sequence of 3D coordinates for the UAV's positions, is denoted by \( \mathbf{q}_i = \{p(n) = (x_n, y_n, h_n) \mid n = 0, 1, \dots, N\} \), where \( p(n) \) refers to the UAV's position at step \( n \), and \( i \) denotes the index of a particular trajectory. The trajectory adheres to specific constraints: horizontal boundaries, altitude limits to avoid collisions and respect operational ceilings, and matching the initial and final positions precisely. Each trajectory \( \mathbf{q}_i \) is evaluated based on its efficiency in reaching the target while maintaining communication quality and minimizing path way.

\vspace{-2mm}
\subsection{Problem Formulation}

We consider a \ac{UAV} as an \ac{O-RU} providing services to users in an urban environment. We optimize the trajectory from the \ac{UAV}'s initial position \( p_s \) to \( p^*_i \). The target location \( p^*_i \) is a dynamic position that changes depending on the distribution of users in target area. We can formulate as:

\begin{equation}\vspace{-3mm}
\mathbf{q}^* = \min_{} \left[ w_1 S(\mathbf{q}_i) + w_2 \Gamma(\mathbf{q}_i) \right]
\end{equation}
subject to: 
\begin{align}
p(0) &= p_s,    \quad p(N) = p^*_i  \\\label{eq:cons3}
|p(n+1) - p(n)| &\leq v_{\max} \Delta t \\  
h_{\min} &\leq h_{\text{UAV}} \leq h_{\max}\\
S(\mathbf{q}_i) &\leq S_{\max}
\vspace{-2mm}
\end{align}
where \( \mathbf{q}_i \) is the trajectory of \ac{UAV}, \( S(\mathbf{q}_i) \) represents the number of steps, corresponding to the distance the \ac{UAV} must travel to reach its destination. \( \Gamma(\mathbf{q}_i) \) represents the number of outage events along the trajectory, indicating disruptions in communication with the ground as the \ac{UAV} travels. Our objective is to minimize \(\Gamma(\mathbf{q}_i) \) and \( S(\mathbf{q}_i) \). The weights \( w_1 \) and \( w_2 \) balance the trade-offs between minimizing the travel distance and avoiding outage areas. Constraint \eqref{eq:cons3} ensures that the distance traveled in a single step cannot exceed the \ac{UAV}'s maximum velocity \( v_{\max} \) and the time step duration \( \Delta t \). The constraint \(S(\mathbf{q}_i)\leq S_{\max} \) ensures that the total number of steps in the trajectory does not exceed a maximum value \( S_{\max} \), which represents the \ac{UAV}'s battery life. We employ a Dueling \ac{DDQN} with multi-step learning to solve this trajectory optimization problem. The Q-function is decomposed as:
\vspace{-2mm}
\begin{equation}
Q(s,a;\theta,\alpha,\beta) = V(s;\theta,\beta) + A(s,a;\theta,\alpha)
\end{equation}
where \( V(s;\theta,\beta) \) is the state value function, and \( A(s,a;\theta,\alpha) \) is the advantage function that represents the relative importance of each action in the given state.

Our primary objective is to minimize both the energy consumption and training time across different environments using \ac{TL}. Given a set of distinct environments \( \{ X_1, X_2, \dots, X_N \} \), we define the \ac{TL} objective as:

\begin{equation}
\theta^* = \arg\min_{\theta} \left[ E(\theta, \theta_{\text{base}}) + T(\theta, \theta_{\text{base}}) \right]
\end{equation}
where \( E(\theta, \theta_{\text{base}}) \) represents the energy consumption of training the model in a new environment, starting from the pre-trained model \( \theta_{\text{base}} \) obtained from a base environment. Similarly, \( T(\theta, \theta_{\text{base}}) \) represents the training time required to adapt the model in the new environment. The goal is to minimize the combined consumption of energy and time when transferring the learned knowledge from the base environment to new environments, thus improving the overall efficiency of the training process.

\section{DRL and Continuous \ac{TL} for UAV Trajectory}

\subsection{Dueling DDQN with Multi steps for UAV Trajectory}

In this research, we employ a Dueling \ac{DDQN} with multi-step learning for optimizing \ac{UAV} trajectories in O-RAN environments. This advanced reinforcement learning technique is particularly well-suited to our problem due to several key features. The dueling architecture separates the estimation of state values and action advantages, allowing a better assessment of \ac{UAV} positioning relative to user demand and network conditions. Double Q-learning reduces overestimation bias, which is crucial for avoiding suboptimal flight paths and energy waste. Multi-step learning enables consideration of the long-term consequences of \ac{UAV} movements, essential in dynamic O-RAN environments. \ac{DDQN} effectively handles the continuous and high-dimensional state spaces created by 3D \ac{UAV} positioning and dynamic user distribution. Its epsilon-greedy strategy with decay balances exploration of new areas with exploitation of known good positions. By leveraging these advantages, our approach enables more efficient and effective \ac{UAV} trajectory design, optimizing coverage and energy efficiency while adapting to the evolving demands of next-generation wireless networks.

In our proposed \ac{UAV}-based O-RAN system, the concept of a target area and the determination of an optimal target location for the \ac{UAV} are crucial components. The target area, denoted as $\mathcal{A}_{\text{target}} \subset \mathbb{R}^2$, is a specific region within the urban environment where service is needed. The target area represents a zone of high user demand or an area requiring enhanced network coverage. The determination of the optimal target location for the UAV, denoted as $p^*$ $(x^*_{\text{UAV}}, y^*_{\text{UAV}}, h^*_{\text{UAV}})$. This optimal position is calculated based on the locations of the $M$ users within $\mathcal{A}_{\text{target}}$. The objective in determining $p^*$ is to maximize the number of users served with satisfactory SINR. 

\subsubsection{MDP for UAV Trajectory}

The \ac{UAV} trajectory optimization problem is formulated as a \ac{MDP}, characterized by a tuple \( (\mathcal{S}, \mathcal{A}, \mathcal{P}, \mathcal{R}, \gamma) \). Each state \( s \in \mathcal{S} \) represents the \ac{UAV}’s current position, which defined as $s = p(t)$. The action space \( a \in \mathcal{A} \) consists of possible movements that \ac{UAV} can make at any given time, accounting for its ability to navigate in multiple directions in 3D space: $\mathcal{A} = \{ \text{forward}, \text{backward}, \text{left}, \text{right} \}$. The state transition function \( \mathcal{P}(s'|s, a) \) describes the probability of transitioning from state \( s \) to state \( s' \) given the action \( a \). In our scenario, the \ac{UAV}'s movement is deterministic, meaning the next state \( s' \) is directly determined by the current state \( s \) and action \( a \). The reward function \( \mathcal{R}(s, a, s') \) is designed to guide the \ac{UAV} in minimizing travel time, avoiding areas with poor signal quality, and maintaining high communication performance. The reward at each step is defined as:
\begin{equation}
\mathcal{R}(s, a, s') = - P_{\text{outbound}} - w_1 \cdot d(s, s_{\text{target}}) - w_2 \cdot \Phi(p(t)) + P_{\text{reach}}
\label{eq:reward}
\end{equation}

where: \( w_1 \) and \( w_2 \) are weight factors that balance the trade offs between reaching the target and avoiding high outage rate areas. \( d(s, s_{\text{target}}) \) is the distance from the \ac{UAV}’s current position to the target position. \( \Phi(p(t)) \) is a penalty applied for passing through regions with poor signal quality, determined by the SINR at position \( p(t) \). \(P_{\text{outbound}}\) is a penalty incurred at each time step to discourage the \ac{UAV} from moving outside the predefined boundaries or taking inefficient steps. \( P_{\text{reach}} \) is a reward given when the \ac{UAV} successfully reaches its destination. The objective is to discover the optimal policy \( \pi^* \) that maximizes the cumulative expected reward across the UAV's entire trajectory, expressed as:
\begin{equation}
\pi^* = \arg\max_{\pi} \mathbb{E} \left[ \sum_{t=0}^{T} \gamma^t \mathcal{R}(s_t, a_t, s_{t+1}) \right],
\end{equation}
where \( \mathbb{E} \) denotes the expected value.

\subsection{\ac{TL} for Training Energy Reduction}

\begin{algorithm}[ht]\small
\caption{Continuous Transfer Learning for UAV Trajectory Optimization in O-RAN}\label{algo:2}
\begin{algorithmic}[1]
\State Initialize base model (Dueling DDQN): weights $\theta_1$, policy $\pi_1$, learning rate $\alpha_1$, exploration $\epsilon_1$
\State Train in Environment 1 (Simulation ENV1), record time $T_1$ and energy $E_1$
\State $N_{\text{environments}} \leftarrow $ ( ENV1, ENV2, Rosslyn, Ottawa )
\For{$i = 1$ to $N_{\text{environments}}$}
    \State $\theta_i \leftarrow \theta_{i-1}$, $\pi_i \leftarrow \pi_{i-1}$ 
    \State Set $\alpha_i$, $\epsilon_i$, reward function $R_i(s, a)$
    \For{episode = 1 to $N_{\text{epi},i}$}
        \State Initialize state $s_0$ in Environment $i$
        \State $E_{\text{episode}} \leftarrow 0$, Initialize episode energy consumption
        \For{$t = 0$ to $N_{\text{step},i}$}
            \State Choose $a_t$ using $\epsilon_i$-greedy policy based on $\pi_i$
            \State Execute $a_t$, observe $s_{t+1}$, get reward $R_i(s_t, a_t)$
            \State $E_{\text{step}} \leftarrow$ MeasureEnergy(), Record energy
            \State $E_{\text{episode}} \leftarrow E_{\text{episode}} + E_{\text{step}}$
            \State Update replay buffer $\mathcal{D}_i$
            \State Update $\theta_i$ using Dueling DDQN
            \State Update target network if necessary
        \EndFor
        \State $E_i \leftarrow E_i + E_{\text{episode}}$ ,Accumulate total energy
    \EndFor
    \State Record $T_i$ (total training time for Environment $i$)
    \State $\eta_{\text{time},i} = T_i / T_1$, $\eta_{\text{energy},i} = E_i / E_1$
    \If{$i = 2$}
        \State Environment $i \leftarrow$ Simulation ENVs
    \ElsIf{$i = 3$}
        \State Environment $i \leftarrow$ Ray tracing ENVs
    \EndIf
\EndFor
\State \textbf{return} $\theta_N$, ${\pi_i}_{i=1}^N$, $\eta_{\text{time},i}$, $\eta_{\text{energy},i}$
\end{algorithmic}
\end{algorithm}

We present a \ac{CTL} approach designed to optimize \ac{UAV} trajectories across multiple O-RAN environments while minimizing training time and energy consumption (Algorithm \ref{algo:2}). Our method leverages knowledge gained from previous environments to accelerate learning in new scenarios, thereby reducing computational overhead and energy expenditure.
The core of our approach is a Dueling \ac{DDQN} that learns to navigate \acp{UAV} carrying \ac{O-RU} in various urban landscapes. By transferring learned weights and policies between environments, we aim to achieve faster convergence and improved energy efficiency in training.

\subsection{Environments}
\begin{figure}[ht]
    \centering
    \begin{minipage}[b]{0.45\columnwidth}
        \centering
        \includegraphics[clip, trim=0cm 0.cm 0.0cm 3cm, width=\columnwidth]{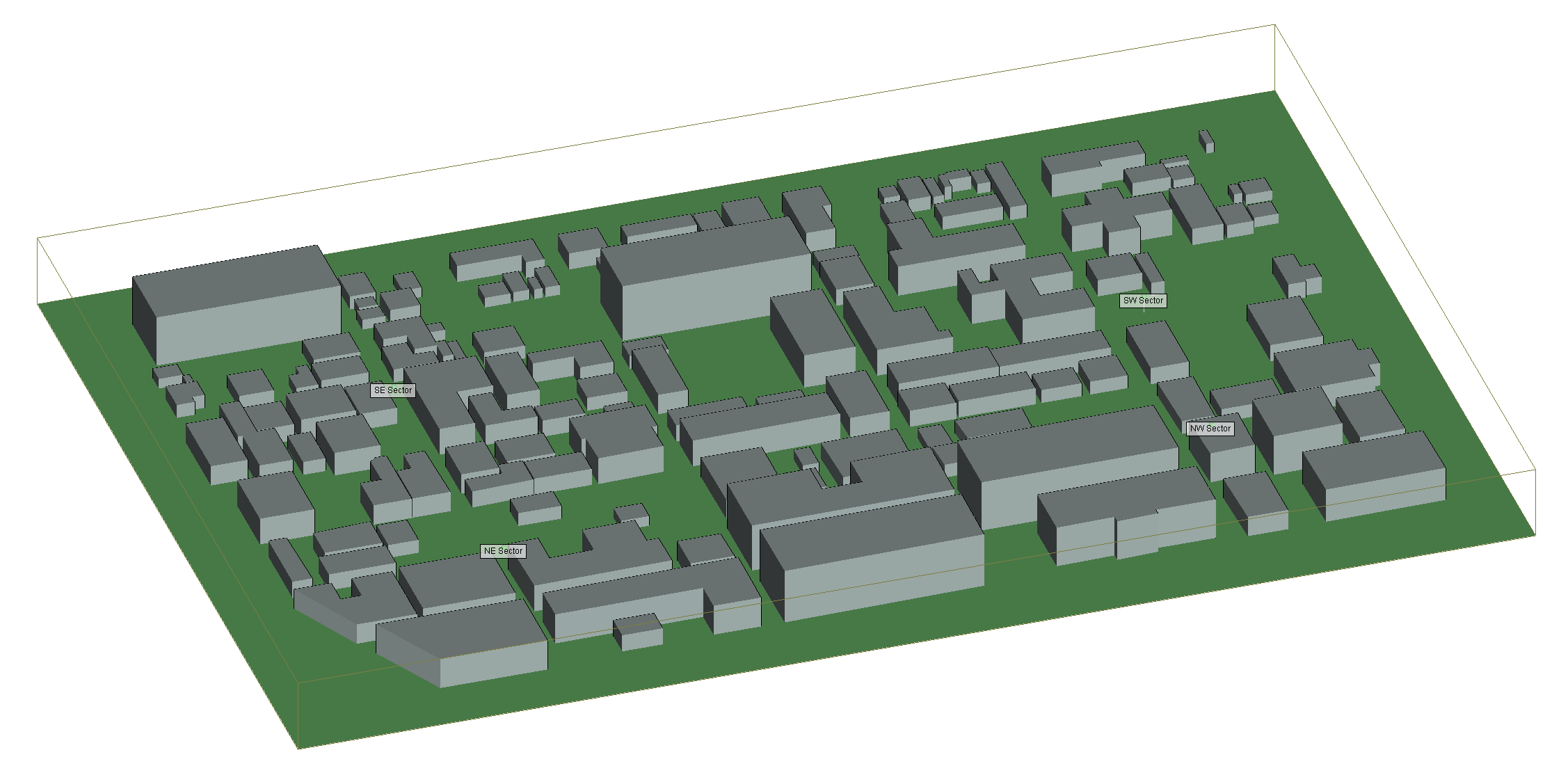}
        \caption{Ottawa city map by Wireless InSite}
        \label{fig:Ottawa}
    \end{minipage}
    \hfill
    \begin{minipage}[b]{0.45\columnwidth}
        \centering
        \includegraphics[clip, trim=0cm 0.cm 0.0cm 3cm, width=\columnwidth]{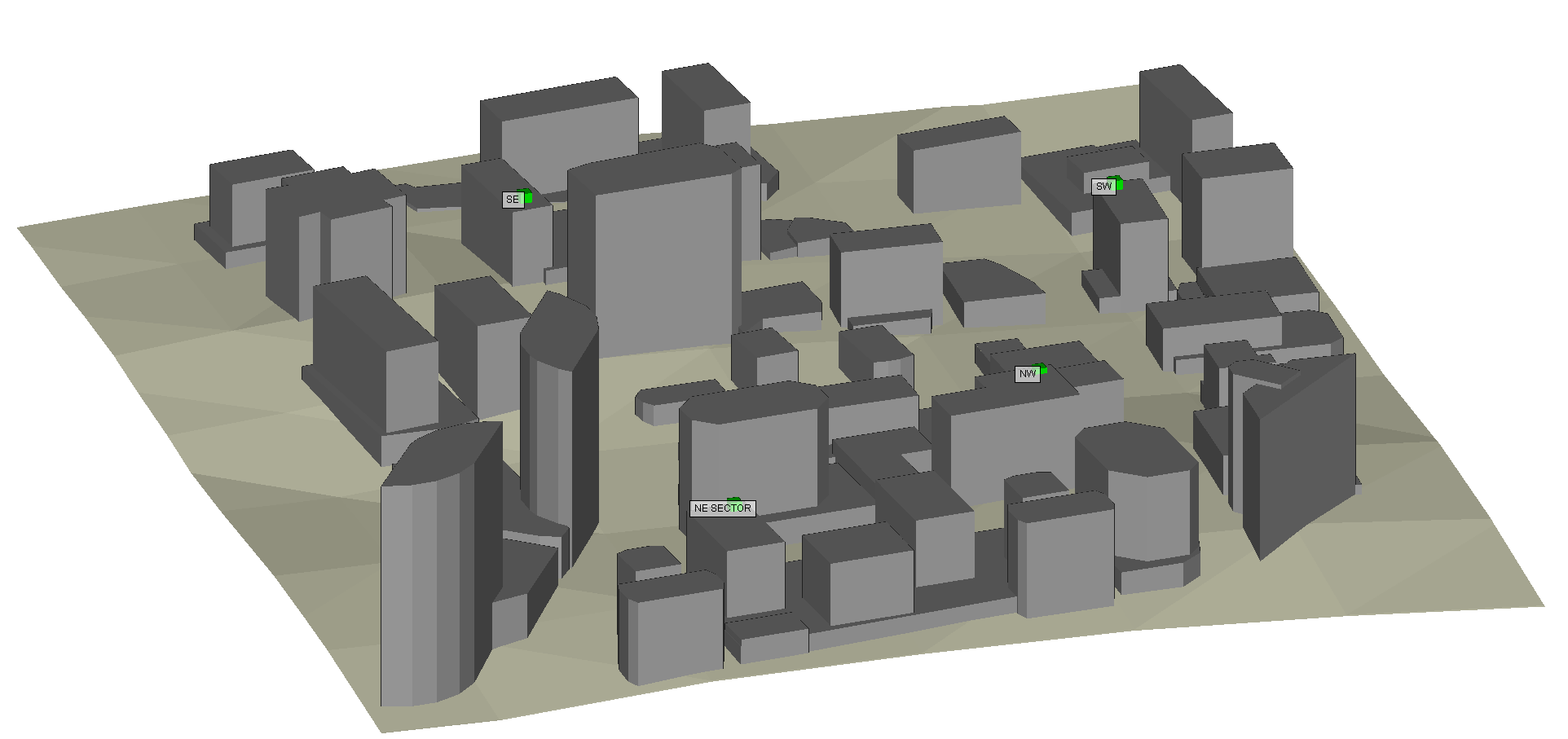}
        \caption{Rosslyn city map  Wireless InSite}
        \label{fig:rosslyn}
    \end{minipage}
    \vspace{-2mm}
\end{figure}
We defined a series of environments to test the adaptability and effectiveness of our continuous \ac{TL} approach. For the simulations, we used Python to model urban landscapes, creating two distinct environments. One of these environments served as the training ground for the Dueling \ac{DDQN} base model, characterized by densely packed buildings in urban areas. The first environment used for \ac{TL} presents a stark contrast, with lower but more densely arranged buildings. This setting tests the \ac{UAV}'s ability to adjust its learning strategy to a drastically different urban topology. The second environment simulates low-rise residential neighborhoods with distinct BS-distribution patterns. Here, the changes in the radio map are more significant, although the target areas are relatively closer. 

\begin{figure}[ht]
\vspace{-2mm}
    \centering
    \includegraphics[width=1\columnwidth]{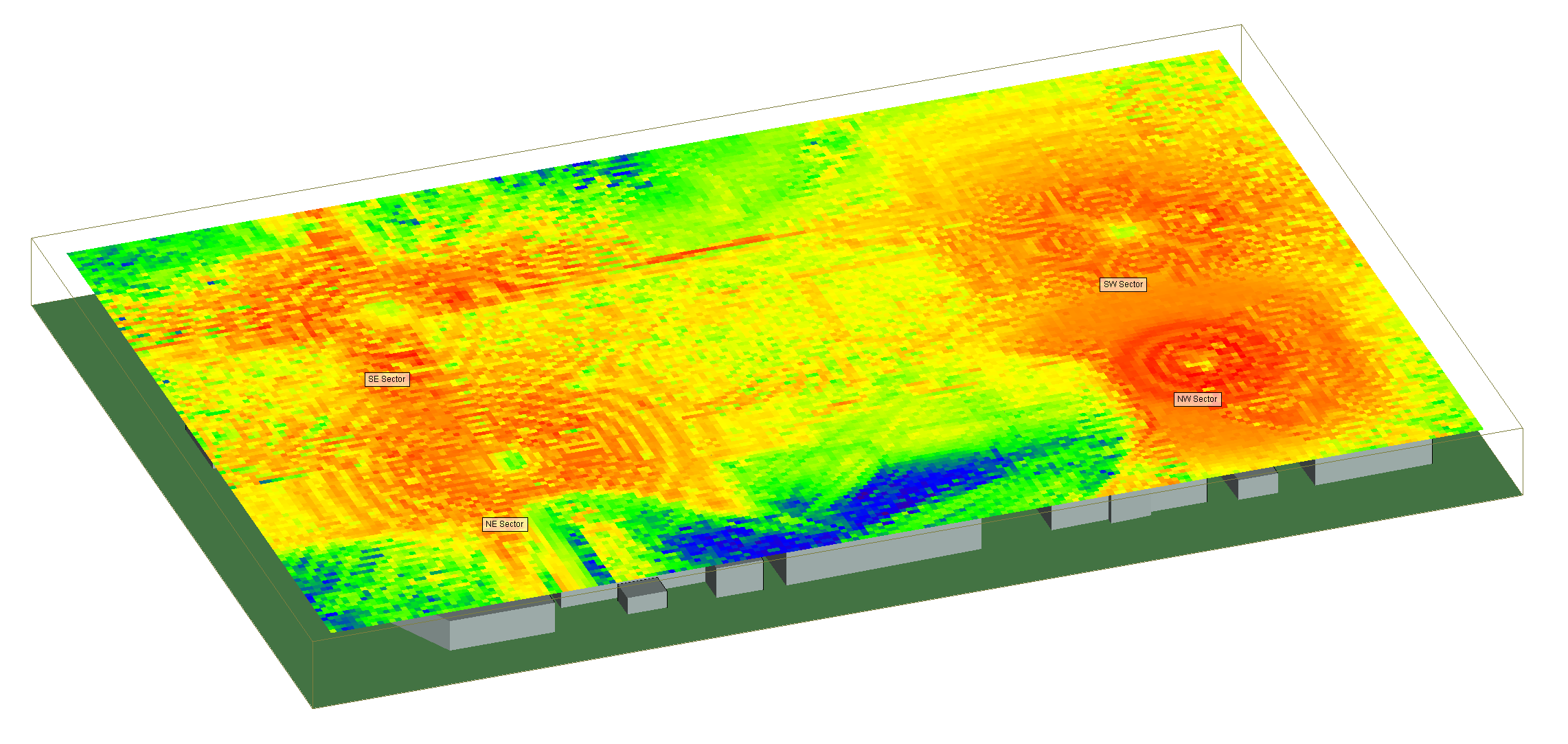}
    \caption{Ottawa radio map of UAV level}
    \label{fig:radiomap}
\vspace{-2mm}
\end{figure}

\begin{figure}[ht]
\vspace{-2mm}
    \centering
    \includegraphics[width=1\columnwidth]{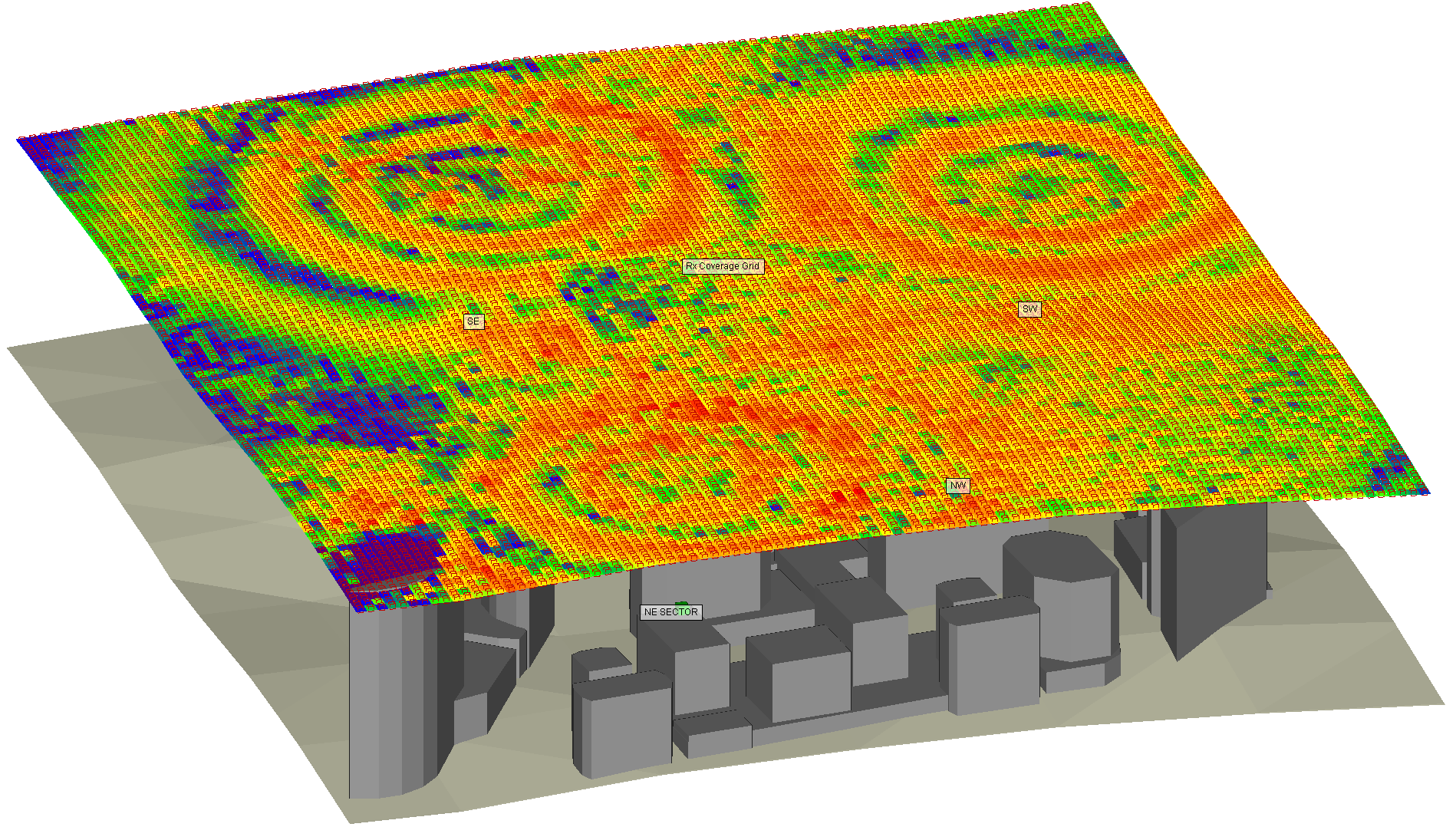}
    \caption{Rosslyn radio map of UAV level}
    \label{fig:radiomap2}
\vspace{-2mm}
\end{figure}

To evaluate the real-world applicability of our approach, we next conducted tests using Wireless InSite \cite{remcom2023wireless} as well, a ray-tracing simulation tool chosen for its ability to deliver highly accurate propagation modeling in complex urban environments. Wireless InSite allows us to import real city maps and detailed 3D building models, enabling high-fidelity simulation of radio wave propagation while accounting for phenomena such as reflection, diffraction, and scattering. Connecting the controlled simulation environments to real-world scenarios is crucial for validating the \ac{TL} approach. While simulations provide an initial testing ground, they often lack the complexity of real-world phenomenas. The use of Wireless InSite provides several key advantages. First, it allows us to validate the performance of our model in realistic, complex environments. Second, it captures detailed multipath effects and material-specific interactions that are critical for accurate signal propagation. Third, it enables us to assess the impact of real-world obstacles and terrain on signal quality and coverage. Finally, it offers a more precise estimation of achievable data rates and network coverage in practical settings. By testing our continuously trained model in these detailed, real-world simulations, we are able to confirm its effectiveness and robustness for practical deployment scenarios. This step ensures that our approach remains applicable and reliable even in challenging, real-world environments. The first map selected for Wireless InSite is of Ottawa, Canada, centered at coordinates longitude 75.667 and latitude 45.3176, which are shown as Fig. \ref{fig:Ottawa}. The area spans 960 m by 630 m and is situated in a densely built urban region. The second map used was of Rosslyn, Arlington, Virginia, located at a latitude of 38.8940 and longitude of -77.0752, near the iconic Rosslyn Twin Towers, which shown as Fig. \ref{fig:rosslyn}. In comparison to Ottawa, Rosslyn features taller buildings and a more complex urban landscape, necessitating higher-altitude UAV communications to navigate the challenging terrain.

These two maps includE material properties for greater accuracy during the simulation, accounting for materials such as dielectric half-space (concrete, wet earth) and one-layer dielectric (brick). For the terrestrial base station, four half-wave dipole antennas are used, transmitting at 30 dBm power. The  \ac{UAV} flying at an altitude of 80 m (Ottawa) and 100 m (Rosslyn). To simulate the communication performance at the  \ac{UAV}'s altitude, we set up receivers in Wireless InSite at \ac{UAV} height using a dense grid, spaced 5 m apart. In the Ottawa map, this resulted in a data grid of 192 by 126 points. However, this grid is not fully compatible with  \ac{UAV} trajectory simulations because the  \ac{UAV} may pass through areas where no communication data is available, leaving gaps in the data. To address this issue, we processed the data by interpolating the missing points, using the median value to fill in the gaps. The processed data was then rescaled to match the full 960 m by 630 m area. Afterward, we converted the SINR data into outage rate values. As shown in Fig. \ref{fig:radiomap}, red areas indicate strong signal strength and low outage rates, while blue areas represent weak signals and high outage rates. Then use the same method to process Rosslyn's map as Fig. \ref{fig:radiomap2} shows.

\section{Results}

\subsection{Transfer learning in Simulation Environment}

To effectively monitor and optimize the energy consumption of the training process~\cite{guerra2023cost}, we utilize the PyJoules toolkit. This allows us to track the power usage of both GPU and CPU in real-time as the machine learning model is trained. The hardware was used in this research is the NVIDIA GeForce RTX 3070 Ti Laptop GPU and the 12th Gen Intel(R) Core(TM) i9-12900H CPU. By understanding the energy footprint of each component, we can gain valuable insights into resource utilization and identify opportunities for optimization. By integrating PyJoules into the training workflow, we can track the energy consumption in a detailed and systematic manner. The tool enables real-time measurements, recording the power usage of both the CPU and GPU, providing an accurate representation of the energy consumption associated with training. This data allows for better energy management and potential reductions in the overall energy footprint of the training process.
 
Our study compared the performance of a \ac{TL} approach against a traditional Dueling \ac{DDQN} with multi-step learning, which is shown in Table \ref{tab:performance}. The \ac{TL} model is derived by adapting the pre-trained model to a new environment and Dueling \ac{DDQN} trained from scratch in the new environment. The \ac{TL} method demonstrated remarkable efficiency in adapting to the new environment. It reached the performance threshold significantly faster than the \ac{DDQN} approach, crossing the threshold at episode 1023, compared to episode 1647 for \ac{DDQN}. This represents a substantial 38\%  reduction in the number of episodes required to achieve the same level of performance.

\begin{table}[ht]
\centering
\setlength{\tabcolsep}{4pt} 
\begin{tabular}{|l|c|c|c|c|}
\hline
\multirow{2}{*}{Metric} & \multicolumn{2}{c|}{Simulation ENV1} & \multicolumn{2}{c|}{Simulation ENV2} \\
\cline{2-5}
 & DDQN & Transfer & DDQN & Transfer \\
\hline
Convergence Episodes & 1950 & 1050 & 2430 & 1350 \\
\hline
Convergence Time (h) & 2.43 & 1.07 & 3.05 & 1.63 \\
\hline
Episodes to 95\% Success Rate & 2100 & 650 & 2300 & 750 \\
\hline
Energy Consumption (Wh) &  2.34 & 1.62 & 2.15 & 0.89 \\
\hline
\end{tabular}
\caption{Performance Comparison of Dueling DDQN with Multi-steps and Transfer Learning in Simulation environment}
\label{tab:performance}
\end{table}

Furthermore, the \ac{TL} approach exhibited superior initial performance. From the outset, it maintained higher reward values, indicating that the knowledge transferred from the first environment provided a significant head start in the new setting. This advantage is particularly valuable in real-world applications where quick adaptation to new scenarios is crucial.
The stability of the \ac{TL} method was also noteworthy. The moving average range for the \ac{TL} approach was narrower than that of the \ac{DDQN}, suggesting more consistent performance throughout the training process. In terms of training energy consumption, both simulation environments, Env1 and Env2, show varying degrees of improvement with the use of \ac{TL}. In Env1, energy consumption is reduced by a modest 30.52\%, indicating that while \ac{TL} is beneficial, certain characteristics of this environment may limit the full potential of the pre-trained knowledge. In contrast, Env2 demonstrates a significant 58.51\% reduction in energy consumption, suggesting that its features align well with the pre-trained model, leading to substantial energy optimization.

\subsection{Transfer learning in Ray tracing environment}

\begin{figure}[ht]
\vspace{-2mm}

    \centering
    \includegraphics[width=0.9\columnwidth]{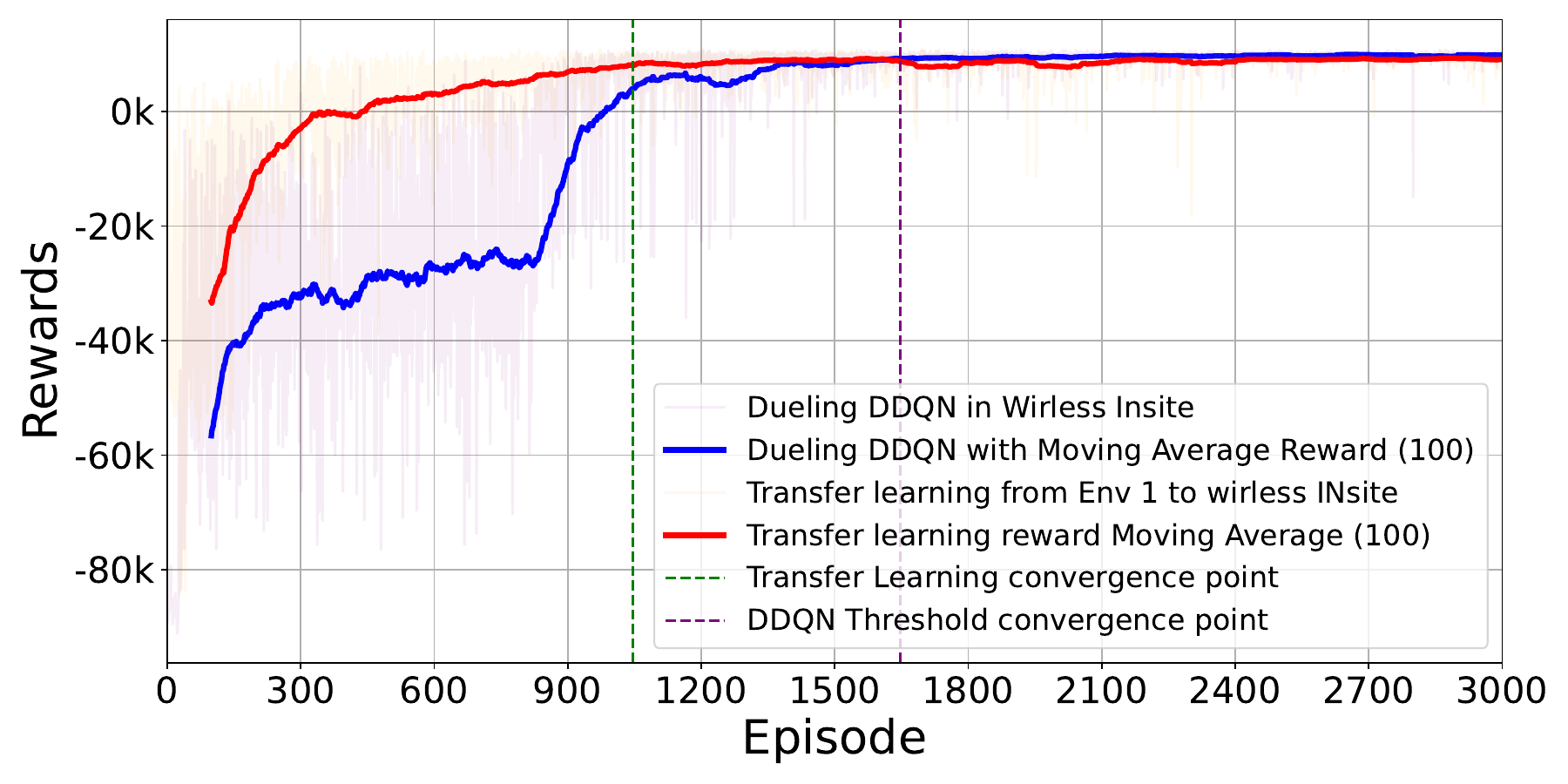}
    \caption{Reward for Wirless Insite in Ottawa}
    \label{fig:1.png}
\vspace{-4mm}
\end{figure}

\begin{figure}[ht]
\vspace{-2mm}

    \centering
    \includegraphics[width=0.9\columnwidth]{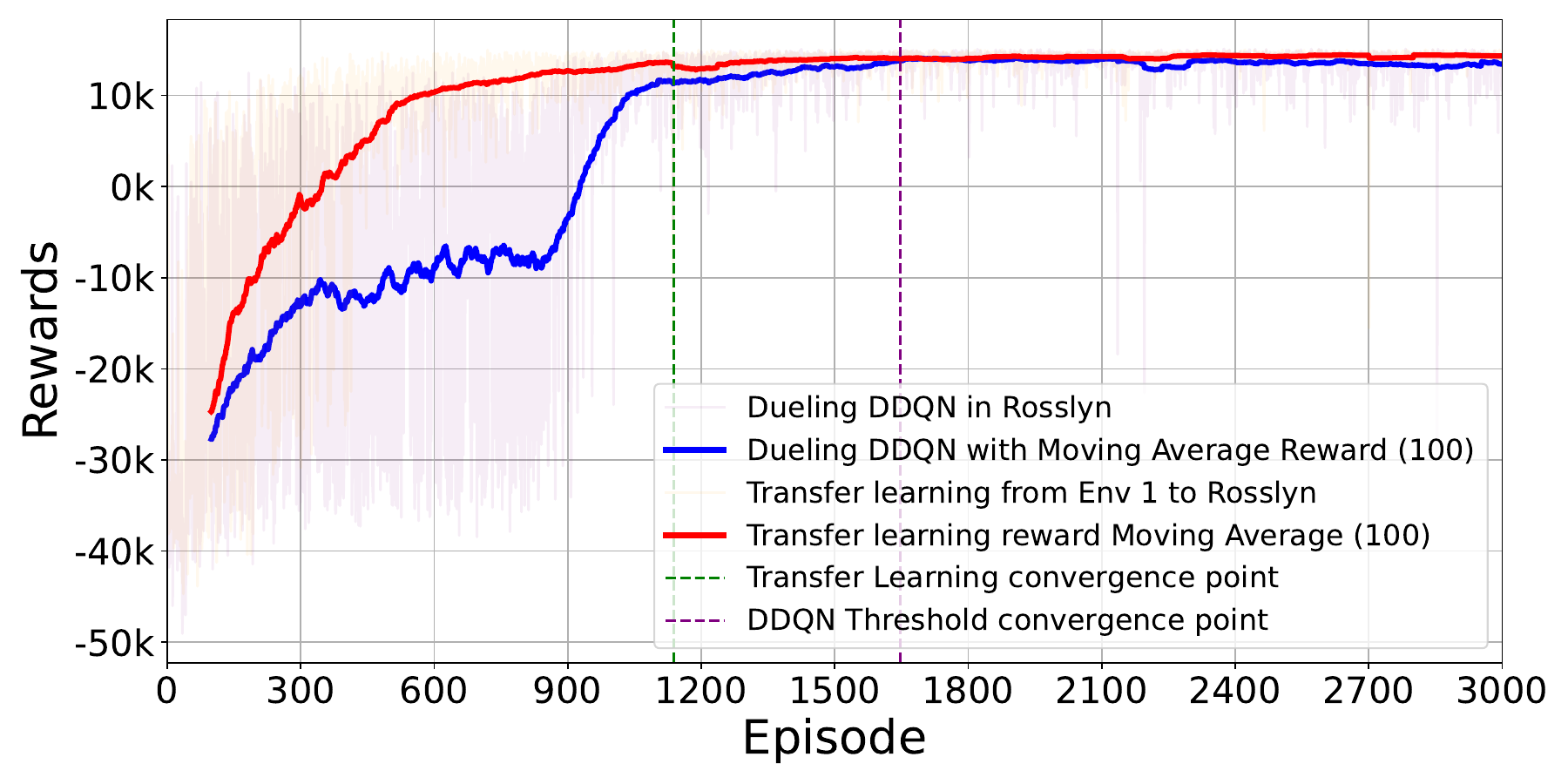}
    \caption{Reward for Wirless Insite in Rosslyn}
    \label{fig:1.png}
\end{figure}

\begin{figure}[ht]
\vspace{-2mm}

    \centering
    \includegraphics[width=0.95\columnwidth]{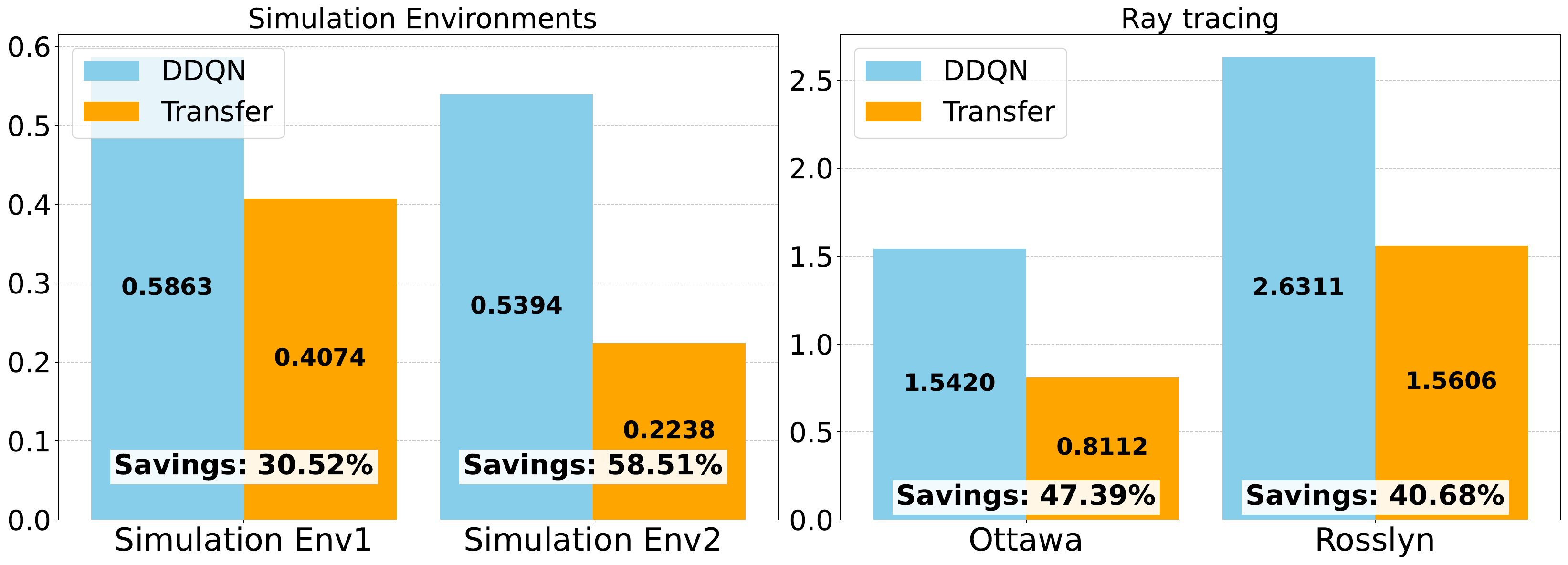}
    \caption{Energy Consumption Comparison: DDQN vs TL. Left represents the simulated environment, right the real city map scenarios}
    \label{fig:2.png}
\vspace{-4mm}

\end{figure}

The test conducted using the Ottawa city model demonstrated significant improvements in both training efficiency and energy consumption. Similar to the simulation environment, the \ac{TL} model is derived from a pre-trained model adapted to a new city, and its performance is compared to that of a Dueling \ac{DDQN} model trained from scratch in the same new city. In Ottawa, the baseline Dueling \ac{DDQN} required approximately 1750 episodes to achieve the desired convergence target; the \ac{TL} approach substantially improved efficiency, requiring only about 1000 episodes. Normalization was used in this analysis to enable a fair comparison of energy consumption across environments of different sizes (Simulation Env1 and Env2: $2000 \times 2000$, Ottawa: $960 \times 630$, Rosslyn: $500 \times 500$). Without normalization, larger environments would naturally show higher total energy consumption, potentially masking efficiency differences. By scaling all energy values to a standard $1000 \times 1000$, the Ottawa environment exhibits substantial improvement with the \ac{TL} approach, achieving 47.39\% energy savings compared to \ac{DDQN}. Similarly, in the Rosslyn environment, the \ac{TL} method resulted in a 36.97\% reduction in training energy, with \ac{TL} achieving convergence at episode 1106, while \ac{DDQN} crossed convergence at episode 1648. Moreover, the \ac{TL} method outperformed \ac{DDQN} in Rosslyn in terms of fewer steps and lower trajectory energy consumption. Rosslyn’s energy consumption is slightly higher for both models due to the increased complexity of the communication channels, stemming from the more complex urban environment, taller buildings, and higher required \ac{UAV} altitudes. Additionally, the \ac{TL} approach has proven its adaptability by working effectively across real-world maps of varying sizes, building heights, and user distributions. This demonstrates its capability to scale across diverse environments with distinct geographic and structural challenges. It is noteworthy that the Ottawa and Rosslyn environments show higher normalized energy consumption for both \ac{DDQN} and \ac{TL} methods compared to the simulated environments, likely due to increased complexity and more challenging environmental factors in real-world settings.

\vspace{-2mm}

\section{Conclusion}

We propose a novel approach where \acp{UAV} functions as mobile \ac{O-RU} within the O-RAN architecture, optimizing their trajectories using Dueling Double DDQN with multi-step learning. This method adapts to dynamic user demand and network conditions efficiently. Our dual-environment testing, using simulations and Wireless Insite ray-tracing with real-world data, validates the system's performance in both synthetic and realistic urban scenarios. By incorporating \ac{TL}, we reduce training time and energy consumption by 30.52\% and 58.51\% in the simulation environment; 44.85\% and 36.97\% in the ray tracing environment, enabling faster adaptation to new environments. This approach balances energy efficiency, training, and trajectory performance in O-RAN \ac{UAV} system. Future work will extend the system to handle more complex, large-scale environments with varying user densities and mobility patterns.

\begin{acronym} 
\acro{5G}{Fifth Generation}
\acro{6G}{Sixth Generation}
\acro{AI}{Artificial Intelligent }
\acro{ACO}{Ant Colony Optimization}
\acro{ANN}{Artificial Neural Network}
\acro{BB}{Base Band}
\acro{BBU}{Base Band Unit}
\acro{BER}{Bit Error Rate}
\acro{BS}{base station}
\acro{BW}{bandwidth}
\acro{C-RAN}{Cloud Radio Access Networks}
\acro{O-CU}{Open Central Unit}
\acro{O-RU}{Open Radio Unit}
\acro{O-DU}{Open Distributed Unit}
\acro{CAPEX}{Capital Expenditure}
\acro{CoMP}{Coordinated Multipoint}
\acro{CR}{Cognitive Radio}
\acro{CRLB}{Cramer-Rao Lower Bound}
\acro{C-RAN}{Cloud Radio Access Network}
\acro{CTL}{Continuous Transfer Learning}
\acro{D2D}{Device-to-Device}
\acro{DAC}{Digital-to-Analog Converter}
\acro{DAS}{Distributed Antenna Systems}
\acro{DBA}{Dynamic Bandwidth Allocation}

\acro{PID}{Proportional–integral–derivative}
\acro{O-RAN}{Open Radio Access Network}

\acro{DC}{Duty Cycle}
\acro{DFRC}{Dual Function Radar Communication}
\acro{DL}{Deep Learning}
\acro{DSA}{Dynamic Spectrum Access}
\acro{DQL}{Deep Q Learning}
\acro{DRL}{Deep Reinforcement Learning}
\acro{DQN}{Deep Q-Network}
\acro{DDQN}{Double Deep Q network}
\acro{DDPG}{Deep Deterministic Policy Gradient}
\acro{FBMC}{Filterbank Multicarrier}
\acro{FEC}{Forward Error Correction}
\acro{FFR}{Fractional Frequency Reuse}
\acro{FL}{Federated Learning}
\acro{FSO}{Free Space Optics}
\acro{FANET}{Flying ad-hoc network}
\acro{GA}{Genetic Algorithms}
\acro{GAN}{Generative Adversarial Networks}
\acro{GMMs}{Gaussian mixture models}
\acro{TF}{Transfer learning}

\acro{KPIs}{Key Performance Indicators}

\acro{Non-RT RIC}{Non-Real Time Radio Intelligent Controller}
\acro{Near-RT RIC}{Near Real-Time Radio Intelligent Controller}

\acro{RIC}{Radio Intelligent Controller}

\acro{HAP}{High Altitude Platform}
\acro{HL}{Higher Layer}
\acro{HARQ}{Hybrid-Automatic Repeat Request}
\acro{HCA}{Hierarchical Cluster Analysis}
\acro{HO}{Handover}
\acro{KNN}{k-nearest neighbors} 
\acro{IoT}{Internet of Things}
\acro{ISAC}{Integrated Sensing and Communication}
\acro{LAN}{Local Area Network}
\acro{LAP}{Low Altitude Platform}
\acro{LL}{Lower Layer}
\acro{LoS}{Line of Sight}
\acro{LTE}{Long Term Evolution}
\acro{LTE-A}{Long Term Evolution Advanced}
\acro{MAC}{Medium Access Control}
\acro{MAP}{Medium Altitude Platform}
\acro{MDP}{Markov Decision Process}
\acro{ML}{Machine Learning}
\acro{MME}{Mobility Management Entity}
\acro{mmWave}{millimeter Wave}
\acro{MIMO}{Multiple Input Multiple Output}
\acro{NFP}{Network Flying Platform}
\acro{NFPs}{Network Flying Platforms}
\acro{NLoS}{Non-Line of Sight}
\acro{RU}{Radio Unit}
\acro{OFDM}{Orthogonal Frequency Division Multiplexing}
\acro{OSA}{Opportunistic Spectrum Access}
\acro{O-RAN}{Open Radio Access Network}
\acro{C-RAN}{cloud radio access network}
\acro{OMC}{O-RAN Management and Control}
\acro{PAM}{Pulse Amplitude Modulation}
\acro{PAPR}{Peak-to-Average Power Ratio}
\acro{PGW}{Packet Gateway}
\acro{PHY}{physical layer}
\acro{PSO}{Particle Swarm Optimization}
\acro{PU}{Primary User}
\acro{QAM}{Quadrature Amplitude Modulation}
\acro{QoE}{Quality of Experience}
\acro{QoS}{Quality of Service}
\acro{QPSK}{Quadrature Phase Shift Keying}
\acro{RF}{Radio Frequency}
\acro{RIS}{Reconfigurable Intelligent Surface}
\acro{RL}{Reinforcement Learning}
\acro{DRL}{Deep Reinforcement Learning}
\acro{RAN}{Radio Access Network}
\acro{RMSE}{Root Mean Squared Error}
\acro{RN}{Remote Node}
\acro{RRH}{Remote Radio Head}
\acro{RRC}{Radio Resource Control}
\acro{RRU}{Remote Radio Unit}
\acro{RSS}{Received Signal Strength}
\acro{SAR}{synthetic-aperture radar}
\acro{SU}{Secondary User}
\acro{SCBS}{Small Cell Base Station}
\acro{SDN}{Software Defined Network}
\acro{SNR}{Signal-to-Noise Ratio}
\acro{SON}{Self-organising Network}
\acro{SVM}{Support Vector Machine}
\acro{TDD}{Time Division Duplex}
\acro{TD-LTE}{Time Division LTE}
\acro{TDM}{Time Division Multiplexing}
\acro{TDMA}{Time Division Multiple Access}
\acro{TL}{Transfer Learning}
\acro{UE}{User Equipment}
\acro{ULA}{Uniform Linear Array}
\acro{UAV}{Unmanned Aerial Vehicle}
\acro{USRP}{Universal Software Radio Platform}
\acro{XAI}{Explainable AI}
\acro{HCP}{heterogeneous computing platform}
 \acro{IoT}{Internet of Things}
\end{acronym}

\vspace{-2mm}

\bibliographystyle{IEEEtran}
\bibliography{references.bib}

\end{document}